\documentclass[aps,prd,preprint,groupedaddress,showpacs]{revtex4-1}
\usepackage{amsmath}
\usepackage{amsfonts}
\def\be{\begin{equation}}
\def\ee{\end{equation}}

\newcommand{\X}{X}
\newcommand{\Y}{Y}
\newcommand{\Z}{Z}
\newcommand{\V}{u}
\newcommand{\U}{w}
\newcommand{\mass}{\rho}
\newcommand{\tension}{\tau}
\newcommand{\flux}{\varphi}

\newcommand{\Lagr}{\mathcal{L}}

\newcommand{\der}[2]{\frac{\partial#1}{\partial#2}}

\newcommand{\tF}{\tilde{F}}
\newcommand{\tN}{\tilde{n}}
\newcommand{\tA}{\tilde{G}}
\newcommand{\tS}{\tilde{\Sigma}}
\newcommand{\non}{\nonumber\\}
\begin{document}

\title{Field Theory for String Fluids}

\author{Daniel Schubring}

\email{schub071@d.umn.edu}

\author{Vitaly Vanchurin}

\email{vvanchur@d.umn.edu}

\affiliation{Department of Physics, University of Minnesota, Duluth, Minnesota, 55812}

\begin{abstract}

We develop a field theory description of non-dissipative string fluids and construct an explicit mapping between field theory degrees of freedom and hydrodynamic variables. The theory generalizes both a perfect particle fluid and pressureless string fluid to what we call a perfect string fluid. Ideal magnetohydrodynamics is shown to be an example of the perfect string fluid whose equations of motion can be obtained from a particular choice of the Lagrangian.  The Lagrangian framework suggests a straightforward extension of the perfect string fluid to more general anisotropic fluids describing higher dimensional branes such as domain walls.  Other modifications of the Lagrangian are discussed which may be useful in describing relativistic superfluids and fluids containing additional currents.

\end{abstract}

\maketitle

\section{ Introduction} 
Many physical, cosmological and biological systems contain extended one dimensional string-like objects. These could be the galactic filaments in the large scale structure, cosmic strings after cosmological phase transition,  fundamental strings near or above the Hagedorn temperature, topological strings in liquid crystals or even more complicated objects such as polymer chains. Most of these systems are considered highly non-perturbative and are usually analyzed using numerical N-body type simulations (see e.g. \cite{Vanchurin:2005yb}). There are, however, a couple of analytical methods that one could potentially adopt to describe a collective behavior of these string-like objects. One possibility is to expand each ``string'' into different vibration modes and to treat these modes as an infinite tower of different ``particles''. Then the problem of strings reduces to the problem of infinitely many particles which can be tackled if the infinite tower is truncated. This is the effective field theory approach taken, for example, by the string theory. 

Another possibility is to first coarse-grain the network of strings and then derive equations of motion for coarse-grained fluids by following the dynamics of microscopically conserved quantities (energy, momentum, tangent vector, etc.) \cite{Schubring:2013qpa}.  This is the hydrodynamic approach that was recently adopted to study, for example, cosmic strings. Note that the two approaches are complementary: the effective field theory is useful when the number of relevant vibration modes is small, and the hydrodynamic description is useful when non-equilibrium effects are suppressed. For example, in the limit of local equilibrium \cite{Vanchurin:2013tk} one can show that the dusts of Nambu-Goto, chiral or, more generally, wiggly strings can be described using the equations for the so-called pressureless string fluid \cite{StringFluid}. In this paper we will start from a (less traditional) hydrodynamic description, but nevertheless seek a (more traditional) field theory description of string fluids.

A field theory Lagrangian for perfect (particle) fluids was known for some time (see e.g. \cite{Schutz:1970my}). More recently a field theory description of perfect fluids with charges was proposed which also allows coupling of fluids to external fields  \cite{DHNS2012}.  In what follows we will describe a simple procedure of constructing field theories of even more general fluids (with pressure, tension, charges, interactions etc.) by considering conserved currents at the level of Lagrangians. Then the hydrodynamic equations of motion can be obtained from the standard variational principle without having to go through a coarse-graining procedure.

The paper is organized as follows. In Sec.\ref{SecPerfectFluid} we define a perfect string fluid as a generalization of an ordinary perfect fluid with an additional conserved flux. The energy-momentum tensor is derived from a Langrangian as a function of three scalar fields. Since the first draft of this paper it was realized that ideal magnetohydrodynamics is a particular example of a perfect string fluid \cite{Dissipative}, and the demonstration of this result is reproduced here.

In Sec.\ref{SecPressureless} we investigate string fluids for which the pressure vanishes. Examples of this case include the Stachel-Letelier model \cite{Stachel}\cite{Letelier1979} and a recent description of coarse-grained Nambu-Goto strings \cite{StringFluid}. It is shown that in these cases the string fluid is foliated by worldsheets of a general form of string described by Brandon Carter \cite{CarterDuality}. And so these classical strings may be alternatively described using the variational principle of this paper.

In Sec.\ref{SecVariational} we discuss the variational principle in more depth, describing the relabeling symmetries of the fields and the corresponding Noether symmetries. And in Sec.\ref{SecClebsch} the relationship of the variational principle to the familiar description of fluids in terms of Clebsch potentials is discussed. By trading one of the previous scalar fields for a Clebsch potential we derive a modified string fluid which is shown to be equivalent to a model of a superfluid by Carter and Langlois \cite{CarterLanglois}.

In Sec.\ref{SecDomainWalls} we extend the perfect string fluid to fluids of higher dimensional branes. A particularly simple case of a fluid foliated by domain walls is discussed and shown to be related to the ordinary theory of a massless scalar field. 

In Sec.\ref{SecCurrents} we extend the perfect string fluid by allowing dependence on additional currents and fluxes such as a conserved entropy density. Two complementary approaches are discussed to achieve this. One approach introduces no extra fields but breaks the relabeling symmetry in the Lagrangian. The other \cite{DHNS2012} maintains the symmetry by introducing additional fields.

In Sec. \ref{SecConclusion} we summarize the main results of the paper and discuss possible future directions of research. In particular we describe how the dissipation effects can be incorporated into presented Lagrangian formulation of string fluids.

\section{Perfect string fluid}\label{SecPerfectFluid} The energy momentum tensor for a perfect fluid is
\begin{align}
T^{\mu\nu} = (\mass + p)\V ^\mu \V ^\nu - p g^{\mu\nu},
\end{align}
where $ \V $ is the unit velocity of the fluid, $ p $ is the pressure, and $ \mass $ is the energy density in the rest frame of $ \V $. The energy density $ \mass$ is a function of the number densities $ n_a $ indexed by $ a $, and we can form the corresponding chemical potentials
\begin{align}
\mu^a \equiv \der{\mass}{n_a}.\label{chem}
\end{align}
These number densities can be the density of any extensive quantity such as baryon number, charge, or entropy (in which case the chemical potential is the temperature). The pressure $p$ is then defined through the usual thermodynamic relation, which defines it essentially as a Legendre transform of $ \mass $,
\begin{align}
p &= -\mass + \mu^a n_a\label{lP}\\
\end{align}
Then using \eqref{chem} we get
\begin{align}
dp &=  - d\mass + \mu_a dn^a+ n_a d\mu^a \non
&= n_a d\mu^a.\label{dP}
\end{align}
In addition to the conservation equation for energy-momentum we have the continuity equations for the currents $ n_a^\mu \equiv n_a \V^\mu$,
\begin{align}
\nabla_\mu n_a^\mu = 0.\label{nCons}
\end{align}
By using these continuity equations and \eqref{dP}, the conservation of $ T^{\mu\nu} $
\[ \nabla_\mu [(\mu^a n_a )\V ^\mu \V ^\nu - p g^{\mu\nu} ]= 0, \]
can be reduced to the following equations of motion:
\begin{align}
n_a^\mu \nabla_{[\mu}(\mu^a u_{\nu]})=0.\label{fluidEq}
\end{align}

What we are calling a perfect string fluid has in addition to the conserved current $ n^\mu $ (we will consider only one current for the moment) a conserved bivector $ F $,
\begin{align}
\nabla_\mu F^{\mu\nu} = 0.\label{Fcons}
\end{align}
This bivector can be understood as representing a conserved flux in the system such as angular momentum or magnetic flux. In the magnetic case $F$ is just the dual of the electromagnetic tensor, and for this reason ideal magnetohydrodynamics can be treated as a special case of this perfect string fluid formalism \cite{Dissipative}. But more generally for any network of oriented strings there will be a bivector associated with the topological flux of strings.

In the perfect string fluid, $ F $ is also constrained to be a \emph{simple} bivector, i.e. it has exactly two linearly independent eigenvectors, and the fluid velocity $ \V $ is in the linear subspace spanned by these eigenvectors. It is then convenient to define the `string flux' scalar $ \flux $ and the normalized bivector $ \Sigma $ as the magnitude and direction of $ F $,
\begin{align}
F^{\mu\nu} = \flux \Sigma^{\mu\nu}\\
\Sigma^{\mu\nu}\Sigma_{\mu\nu}=-2.
\end{align}
The orthogonal to $\V$ spacelike direction $ \U $ is defined from $ \Sigma $ and $ \V $,
\begin{align}
\U^\mu \equiv \Sigma^{\mu\nu}\V_{\nu},\label{defU}
\end{align}
in terms of which we can choose to express $ \Sigma $ as,
\begin{align}
\Sigma^{\mu\nu} = \U^\mu \V^\nu - \V^\mu \U^\nu\\
\V^\mu \V_\mu = - \U^\mu \U_\mu = 1\\
\V^\mu \U_\mu = 0
\end{align}
The projector $ h $ onto the linear subspace spanned by $ \V $ and $ \U $ can also be defined in terms of $ \Sigma $,
\begin{align}
h^{\mu\nu}\equiv \V^\mu \V^\nu - \U^\mu \U^\nu = \Sigma^{\mu\rho}\Sigma_{\rho}^{\,\,\,\nu}.\label{h}
\end{align}

The conservation condition on $ F $ \eqref{Fcons} implies through the Frobenius theorem that $ \V $ and $ \U $ lie along two-dimensional integrable submanifolds that can be identified as string worldsheets \cite{StringFluid}. And the dual tensor to $ F $, $ \tF $ is a two-form that can be integrated to give the flux of these strings across a surface. The conservation of $ F $ just implies that the net flux of strings through any closed surface is zero.

The dual to the current $ n $, which we denote by $ \tN $, will also be useful as it is a three-form that can be integrated over a volume to give the conserved charge contained. These two differential forms, $\tN$ and $\tF$,  have a natural interpretation in terms of Lagrangian coordinates labeling fluid particles. There is a two-dimensional space of distinct `worldsheet' submanifolds that we can label with the coordinates $ \X $ and $ \Y $. There is an implicit map that specifies which worldsheet passes through a given spacetime point that allows us to define $ \X $ and $ \Y $ as functions on spacetime. The two-dimensional surfaces along which both $ X $ and $ Y $ take constant values are just the worldsheets. As we will discuss later there is a great deal of symmetry in how we choose these coordinates but we do choose them so that the measure $ d\X\wedge d\Y $ is just the string flux. In fact this will be taken as a definition,
\begin{align}
\tF \equiv d\X\wedge d\Y,
\end{align}
and thus we define the dual bivector $ F $ in \eqref{Fcons} ultimately in terms of $ X $ and $ Y $ fields.

$X$ and $Y$ specify a distinct worldsheet, but to label the distinct fluid particles along the string we need a third coordinate $ \Z $. The one-dimensional spaces along which all three coordinates are constant are just the particle worldlines. Again we will fix the measure $ d\X\wedge d\Y \wedge d\Z $, taking it to be the number density,
\begin{align}
\tN \equiv d\X\wedge d\Y\wedge d\Z ,
\end{align}
and so the current $ n^\mu $ and thus the directions $ \V $ and $ \U $ (through \eqref{defU}) are also specified in terms of these three scalar fields (the Lagrangian coordinates).

An important thing to note about the use of Lagrangian coordinates is that the continuity equations \eqref{nCons} and \eqref{Fcons} are satisfied by construction,
\begin{align}
d\tN = 0\\
d\tF = 0.\label{dF}
\end{align}
To get a complete set of equations of motion we only need to add the conservation of the energy-momentum tensor which is specified by choosing a Lagrangian as a certain function of the $ \X,\Y$ and $\Z $ fields,
\begin{align}
\Lagr =\Lagr\left (\frac{1}{2} \left (dX \wedge dY \right)^2, -\frac{1}{3!} \left (dX \wedge dY \wedge dZ \right )^2 \right).
\end{align}
Note that the Lagrangian for perfect string fluids, $ \Lagr(\flux,n) $, only depends on the scalar fields through the combinations $ \flux^2 $ and $ n^2 $,
\begin{align}
\flux^2 &= \frac{1}{2}\tF^{\lambda\mu}\tF_{\lambda\mu} = \frac{1}{2} \left (dX \wedge dY \right)^2 \\
n^2 &= -\frac{1}{3!}\tN^{\lambda\mu\nu}\tN_{\lambda\mu\nu} =-\frac{1}{3!} \left (dX \wedge dY \wedge dZ \right )^2 .
\end{align} 
Varying the Lagrangian by $ g_{\mu\nu} $ we find $ T^{\mu\nu} $: 
\begin{align}
T^{\mu\nu} &= 2\der{\Lagr}{g_{\mu\nu}}-\Lagr g^{\mu\nu}\non
&=2[\der{\Lagr}{\flux^2}\flux^2 (g^{\mu\nu}-h^{\mu\nu}) + \der{\Lagr}{n^2}n^2 (g^{\mu\nu}-\V^{\mu}\V^{\nu})]-\Lagr g^{\mu\nu}\non
&=(\mass+p)\V^{\mu}\V^{\nu}-(\tension+p)\U^{\mu}\U^{\nu}-pg^{\mu\nu}\label{T}
\end{align}
where we define
\begin{align}
\mass &\equiv -\Lagr\\
p &\equiv \Lagr  - \Lagr_{,\flux}\flux-\Lagr_{,n}n,\label{p}
\end{align}
and the new thermodynamic potential $ \tension $ related to the string tension,
\begin{align}
\tension \equiv    - \Lagr +\Lagr_{,n}n,\label{tau} .
\end{align}
Energy-momentum tensors of this form have been applied for instance to the study of blackfolds \cite{Blackfold} and anisotropic cosmological models \cite{Letelier1983}\cite{StringCosmology1990}\cite{Wang2002}. Our focus here is to study the variational principle underlying this fluid, and show how modifications of the Lagrangian can lead to more general models of interacting fluids.

First of all, if $ \Lagr $ does not depend on $ \flux $ the perfect string fluid reduces to the ordinary perfect fluid. This approach to perfect fluids in terms of a variational principle and Lagrangian coordinates is well established (see \cite{FluidReview} for a review). Usually the variational principle is expressed by varying the worldlines in the action through diffeomorphisms. But as we show later, we can also treat $ \X,\Y,\Z $ as ordinary scalar fields which can be varied independently to produce the equations of motion.  A similar field theory perspective for perfect fluids is found in \cite{DHNS2012}.

A perfect string fluid can also be understood as a generalization of ideal magnetohydrodynamics \cite{Dissipative}. In the isentropic case the electric field vanishes in the rest frame of the fluid \cite{Eckart1940}. In terms of the electromagnetic field tensor $\tF$ this can be written
\begin{align}
\tF^{\mu\nu}u_\nu = 0.\label{eVanish}
\end{align}
In the string fluid context this is the condition for $F$ to be a simple bivector, and for $u$ to be in its linear subspace. As discussed previously, this implies that there are two dimensional `worldsheets' which are everywhere tangent to the velocity and the magnetic field, which is also in the linear subspace of $F$. But in the context of magnetohydrodynamics this is just the statement that the magnetic field lines are "frozen-in" and dragged along by the velocity of the fluid.

In the standard covariant description of magnetohydrodynamics the energy-momentum tensor is simply the sum of a perfect fluid part and an electromagnetic field part \cite{Harris1957}. Any interaction between the two sectors takes place implicitly through the conservation of energy-momentum and in the frozen-in field line condition. Given the variational principle for a perfect fluid discussed above, the Lagrangian for ideal magnetohydrodynamics can be expressed as
\begin{align}
\Lagr&=-\rho_0(n^2) -\frac{1}{2}\flux^2=-\rho_0\left(\frac{1}{3!} \tilde{n}^{\rho\sigma\kappa}\tilde{n}_{\rho\sigma\kappa}\right) -\frac{1}{4}\tF^{\rho\sigma}\tF_{\rho\sigma}=\\
&=-\rho_0\left ( -\frac{1}{3!} (dX \wedge dY \wedge dZ)^2 \right) - \frac{1}{4}(dX \wedge dY)^2,
\end{align}
where $\rho_0(n^2)$ is the energy density of the perfect fluid as a function of $\tN$.

This leads to an energy-momentum tensor
\begin{align}
T^{\mu\nu}= (\rho_0 + p_0 + \flux^2)u^\mu u^\nu - \flux^2 w^\mu w^\nu -(p_0+\frac{1}{2}\flux^2)g^{\mu\nu},
\end{align}
where $p_0$ is the pressure of the perfect fluid component, which differs from the full string fluid pressure appearing as the coefficient of $g^{\mu\nu}$. The energy-momentum tensor for magnetohydrodynamics has been previously written in this form (e.g. \cite{Ciubotariu1973}). We wish to emphasize that the variational principle for string fluids in terms of scalar fields can be applied to magnetohydrodynamics as well.

\section{Pressureless string fluid}\label{SecPressureless}

Besides the reduction to the perfect fluid, another simplification of the string fluid occurs when  $ \Lagr $ only depends on $\flux$ and not $ n $, a case previously studied by Kopczynski.\cite{Kopczynski} The inspiration behind the Kopczynski fluid came from the case where the pressure vanishes, in which case the string fluid further reduces to a model studied by Stachel in which the submanifolds behave as independent Nambu-Goto strings.\cite{Stachel} More recently it was shown that coarse-graining an interacting network of Nambu-Goto strings in the limit of local equilibrium leads to a pressureless string fluid where the submanifolds behave as wiggly strings\cite{StringFluid}.

 To gain a better understanding of the connection between a pressureless fluid and classical strings, first note that $ \tN = \tF\wedge d\Z $ involves a factor of $ \flux $ and so it may be helpful to define a factored number density $ \nu $,
\begin{align}
n \equiv \flux \nu
\end{align}
From \eqref{p}, the condition for the pressure to vanish is
\begin{align}
\Lagr &= \flux \left(\der{\Lagr}{\flux}\right)_n +n \left(\der{\Lagr}{n}\right)_\flux\non
&=\flux \left(\der{\Lagr}{\flux}\right)_\nu,\nonumber
\end{align}
which implies that the derivative $\Lagr_{,\flux}$ at constant $ \nu $ is a function of $\nu$ alone, which we write as
\begin{align}
\Lagr \equiv - \flux \,U(\nu).\label{Un}
\end{align}
Similarly we can define a modified tension of the same form as \eqref{tau},
\begin{align}
T &\equiv U - U_{,\nu} \nu = \flux^{-1} \tension,
\label{eq_state}
\end{align}
so that the energy momentum tensor is just
\begin{align}
T^{\mu\nu} = \flux (U\V^\mu \V^\nu - T \U^\mu \U^\nu).\label{T_pressureless}
\end{align}
This notation is intentionally similar to that used by Carter in describing ``barotropic'' classical strings \cite{CarterBrane}. The difference is that Carter's formalism applies to a single string rather than a fluid foliated by worldsheets, and so any spacetime derivatives must be projected into the worldsheet directions. For instance, in Carter's formalism the condition for the simple bivector $\Sigma$ to describe an integrable submanifold is given as,
\begin{align}
h^\lambda_{\,\,\,\mu}\nabla_\lambda \Sigma^{\mu\nu} = 0.\label{carter_fro}
\end{align}
This condition corresponds to our conservation of the $F$ tensor
\begin{align}
\nabla_\mu(\flux \Sigma^{\mu\nu})=0.\label{carter_fro_2}
\end{align}
In general it can be proven (using both \eqref{carter_fro} and \eqref{carter_fro_2}) that if there is a tensor $ A^{\mu\dots} $ where the index $ \mu $ lies in the worldsheet, then the following statements are equivalent:
\begin{align}
\nabla_\mu(\flux A^{\mu\dots})=0\non
h^\lambda_{\,\,\,\mu}\nabla_\lambda A^{\mu\dots}=0.\label{carter_correspondence}
\end{align}

Since in the pressureless case the conservation of $ T^{\mu\nu},  F^{\mu\nu},$ and $n^{\mu}$ are all of this form, we see that $ \flux $ decouples from the equations of motion for the submanifolds themselves, and these latter equations only depend on derivatives along the worldsheets. So the motion of each individual submanifold may be solved for independently as a barotropic string described by the equation of state $ U(\nu) $. Once $ \Sigma $ has been solved for, the string flux $ \flux $ is determined by the initial values on any two-dimensional spacelike surface intersecting the submanifolds.

The connection between this variational approach and that of barotropic strings can be used to construct Lagrangians describing a theory with submanifolds acting as strings with an arbitrary equation of state. The simplest case would be trivial equation of state for Nambu-Goto strings where $U=\mu_0$, the constant string tension. The string fluid with submanifolds acting as Nambu-Goto strings is the Stachel model, and from \eqref{Un} we see that the Lagrangian is just
\begin{align}
\Lagr = -\mu_0\flux = -\mu_0\sqrt{(dX\wedge dY)^2}.\label{stachel_L}
\end{align}

A slightly more complicated example would be a fluid description of a system of many Nambu-Goto strings \cite{StringFluid} which has submanifolds behaving as wiggly strings satisfying the condition $ UT = \mu_0^2 $. By \eqref{eq_state}, this is described by the equation of state $ U(\nu)= \mu_0\sqrt{1+\nu^2} $. So the Lagrangian for that model is given by
\begin{align}
\Lagr &= -\mu_0\flux \sqrt{1+\nu^2} = -\mu_0\sqrt{\flux^2+n^2} =\\
&= -\mu_0 \sqrt{\frac{1}{2}(dX\wedge dY)^2-\frac{1}{3!}(dX\wedge dY\wedge dZ)^2},
\end{align}
where again $ \flux^2 $ and $ n^2 $ were rewritten in terms of the scalar fields $ \X,\Y,\Z $.

\section{Variational principle}\label{SecVariational} 

Until now the Lagrangian $ \Lagr $ has only been used to find the energy-momentum tensor. The conservation of $ T^{\mu\nu} $ and the identities $ d\tN = d\tF = 0 $ are all that is needed for the equations of motion, but it is not clear that this is equivalent to requiring that the action $S$ be invariant under variations of $ \X,\Y,\Z $. Writing the action explicitly in terms of these fields,
\begin{align}
S&= \int dx^4 \sqrt{-g}\, \Lagr\left(\,\flux^2,n^2\right)\non
&=\int dx^4 \sqrt{-g}\, \Lagr\left(\,\frac{1}{2}(dX\wedge dY)^2,-\frac{1}{3!}(dX\wedge dY\wedge dZ)^2\right).\label{action}
\end{align}
Since $Z$ only appears in terms of its derivative, the field equation resulting from a variation $\delta Z$ can be expressed as the conservation of a current $\Pi_Z$
\begin{align}
\nabla_\nu (2\der{\Lagr}{n^2}\tN^{\lambda\mu\nu}X_{,\lambda} Y_{,\mu})\equiv \nabla_\nu \Pi_Z^\nu=0.
\end{align}
This can also be understood as the Noether current associated with translations in $ \Z $. Recalling the definition \eqref{defU} of $ \U $ and that of the chemical potential $\mu$ \eqref{chem},
\begin{align}
\Pi_Z^\nu = \flux \mu \U^\nu.\label{dual_current}
\end{align}
Due to the decoupling of $ \flux $ through \eqref{carter_correspondence}, for a pressureless fluid this is identical to the spacelike current that appears as a dual to $n$ in Carter's work on classical strings (e.g. \cite{CarterDuality}). Here we see the connection to translation symmetry of $Z$, and see that an analogue also holds for string fluids with pressure.

The field equation corresponding to a variation $ \delta\X $ can be written as
\begin{align}
X_{,\kappa}Y_{,\mu}\nabla_\lambda \left (\frac{\partial \Lagr}{\partial \flux} \tS^{\lambda\mu} \right )-X_{,\kappa}Y_{[,\mu}Z_{,\nu]}\nabla_\lambda\left (\frac{\partial \Lagr}{\partial n}\tilde{u}^{\lambda\mu\nu} \right )=0.\label{fluidEq_string0}
\end{align}
Putting the $ \delta\Y $ and $ \delta\Z $ equations in the same form and combining leads ultimately to the field equations
\begin{align}
-\frac{3}{2}F^{\lambda\mu}\nabla_{[\kappa}\left (\frac{\partial \Lagr}{\partial \flux} \Sigma_{\lambda\mu]} \right)+2 n^\lambda \nabla_{[\kappa}\left (  \frac{\partial \Lagr}{\partial n} u_{\lambda]} \right )=0.\label{fluidEq_string}
\end{align}
For an ordinary particle fluid the first term vanishes and the second term is just the usual equations of motion \eqref{fluidEq}. On the other hand the first term by itself appears also in Kopczynski's work.\cite{Kopczynski} These field equations can be shown to be equivalent to conservation of $ T^{\mu\nu} $ by reversing the steps leading to \eqref{fluidEq}. Although here we are considering a single current and a single string flux, adding additional dependences $n_a$ and $F_b$ in the Lagrangian simply leads to equations of the same form with a sum over the indices $a,b$. 

Of course just as for $\Pi_Z$, the field equations for $\delta X$ and $\delta Y$ can be understood as the conservation of the Noether currents $\Pi_X, \Pi_Y$ associated with translations in $X,Y$. These are special cases of a larger group of symmetry transformations leaving the two-form $dX\wedge dY$ invariant. This group is equivalent to the symplectic transformations on the two-dimensional $X,Y$ space. A symplectic transformation can be generated by an arbitrary function $H(X,Y)$, where for infinitessimal $\delta t$
\begin{align}
\delta\X &= +H_{,Y}\delta t\non
\delta\Y &= -H_{,X}\delta t.\label{symplectic}
\end{align}
Symmetry under these relabeling transformations corresponds to the conservation of the class of currents
\begin{align}
\nabla_\mu(H_{,Y}\Pi_X^\mu - H_{,X}\Pi_Y^\mu) =0,\label{noether}
\end{align}
which is in turn equivalent to certain conditions on $\Pi_X,\Pi_Y$,
\begin{align}
\Pi_X^{\mu} Y_{,\mu} =\Pi_Y^\mu X_{,\mu} = 0\non
\Pi_X^\mu X_{,\mu} =\Pi_Y^\mu Y_{,\mu}.
\end{align}
Similarly an arbitrary function of $X,Y$ may be added to $Z$ without changing the physical situation, and the conservation of the corresponding Noether currents is equivalent to the condition
\begin{align}
\perp^{\lambda}_{\,\,\mu}\Pi_Z^{\mu}= 0.\label{noether_perp}
\end{align}
where
\begin{align}
\perp^{\lambda}_{\phantom{\lambda}\mu} = \delta^\lambda_{\phantom{\lambda}\mu} - \V^\lambda \V_\mu+\U^\lambda \U_\mu\label{noether_perp}
\end{align}
is the orthogonal projection to `worldsheets' spanned by $\V$ and $\U$.
And so any field theory with the same relabeling symmetries (which may depend on higher order derivatives of the scalar fields) will have field equations equivalent to the conservation of three currents $\Pi_X, \Pi_Y, \Pi_Z$ satisfying the constraints above. 

\section {Domain Wall Fluid}\label{SecDomainWalls}

The variational approach discussed in this paper can be easily generalized to different dimensions of spacetime and to different numbers of scalar fields. In such cases the submanifolds in the fluid may describe the world-volumes of higher dimensional branes instead of strings or particles. A simple case we will treat here is that of a single scalar field $X$ in $3+1$ dimensional spacetime. In this case the $2+1$ dimensional submanifolds along which $X$ is constant can describe the world-volume of two dimensional membranes or domain walls.

The gradient one-form $\tA_\mu \equiv X_{,\mu}$ annihilates the tangent vectors to the world-volume, so the orthogonal projector can be written as,
\begin{align}
\perp_{\mu\nu} = -\frac{1}{\psi^2}X_{,\mu}X_{,\nu},
\end{align}
where $\psi$ is the magnitude of $X_{,\mu}$,
\begin{align}
\psi^2 \equiv -\tA^\mu \tA_\mu.
\end{align}
This quantity $\psi$ can be understood as the density of domain walls along their normal direction, and similarly to $\flux$ and $n$ it may appear in the Lagrangian. As before, the divergence of the dual 3-form to $\tA$ vanishes
\begin{align}
G^{\lambda\mu\nu}\equiv \epsilon^{\lambda\mu\nu\rho}\tA_\rho\non
\nabla_\lambda  G^{\lambda\mu\nu} =0,
\end{align}
and separating $G$ into its magnitude and direction
\begin{align}
G^{\lambda\mu\nu}\equiv \psi \Sigma^{\lambda\mu\nu},
\end{align}
the projector $h$ onto the tangent space of the world-volume can be written
\begin{align}
h^{\mu}_{\,\,\,\nu} = \frac{1}{2} \Sigma^{\mu\rho\sigma}\Sigma_{\nu\rho\sigma}.
\end{align}

Now considering the Lagrangian corresponding to the Stachel model \eqref{stachel_L}
\begin{align}
\Lagr =  -\psi =-\sqrt{-g^{\mu\nu}X_{,\mu}X_{,\nu}} ,
\end{align}
the energy-momentum tensor is
\begin{align}
T_{\mu\nu} &= \frac{1}{\psi}X_{,\mu}X_{,\nu} -\Lagr g_{\mu\nu}\non
&= \psi(-\perp_{\mu\nu}+g_{\mu\nu}) = \psi h_{\mu\nu}.
\end{align}
Expressing $h$ in terms of $\Sigma$, the conservation of energy-momentum leads to
\begin{align}
\Sigma^{\mu\rho\sigma}\nabla_\mu \Sigma_{\nu\rho\sigma}=0,
\end{align}

This equation is the analogue of the equation
$\Sigma^{\mu\rho}\nabla_\mu \Sigma_{\nu\rho}=0$
appearing in the Stachel model of a string fluid \cite{Stachel}. And following exactly the same line of reasoning as in that paper we can choose three coordinates parametrizing the world-volume and define  the maps $\xi^{\mu}$ embedding the world-volume in spacetime. Then $\Sigma$ may be expressed in terms of $\xi$ and ultimately we find
\begin{align}
\xi_{,a}^\mu\nabla_\mu(\sqrt{-h} \xi_{\nu}^{,a})=0,
\end{align}
where $h$ is now the determinant of the projector in the world-volume basis (i.e. it is the determinant of the pullback of the metric). This has exactly the same form as the Nambu-Goto equations of motion, except that $a$ ranges over three coordinates on the world volume rather than two. And these are indeed the standard equations for a domain wall in the limit of zero thickness, see for instance \cite{StringReview}.

As an aside, note that it is easy to also consider the Hamiltonian formulation of this theory of domain wall submanifolds. The conjugate momentum $P$ to $X$ is just the time component of the Noether current $\Pi_X^{\mu}$ associated to translations in $X$,
\begin{align}
P &= \Pi_X^0\non
\Pi_X^\mu &= \der{\Lagr}{X_{,\mu}}=\frac{1}{\psi}g^{\mu\nu}X_{,\nu},\label{PiX}
\end{align}
where, specializing to a Minkowski metric,
\begin{align}
\psi^2 &= \frac{\left|\nabla X\right|^2}{1+P^2}.
\end{align}
Then the Hamiltonian density is found to be
\begin{align}
\mathcal{H} &=  \dot{X} P + \psi \non
& = \psi P^2 +\psi \non
&= \left|\nabla X\right|\sqrt{ 1+P^2 },
\end{align}
and Hamilton's equations just express the conservation of the current $\Pi_X$ and its relation to the time derivative $X_{,0}$. This Hamiltonian may also be of interest in lower spacetime dimensions, where it describes a dust of Nambu-Goto strings or free particles.

Returning now to the Lagrangian formulation and a general metric, we can generalize the Lagrangian to an arbitrary function of $\psi$ and that will lead to the appearance of non-vanishing pressure in the energy-momentum tensor just as in the string fluid case. So in the familiar theory of a massless scalar field $\Lagr = g^{\mu\nu}X_{,\mu}X_{,\nu}$
(where $X_{,\mu}$ is spacelike) the surfaces along which $X$ is constant act like domain walls under pressure.

Another direction of generalization is to reintroduce dependence of the Lagrangian on $n$ in addition to $\psi$ with both quantities expressed in terms of the same scalar field $X$. By construction the fluid velocity described by $\tN$ will be confined to the domain walls, just as before it was confined to the worldsheet submanifolds. In the same way, reintroducing dependence on $\flux$ in the Lagrangian will describe a string fluid confined to domain walls. This can be interpreted in a less obscure way as the Lagrangian for a perfect anisotropic fluid with a distinct pressure (or tension) in three characteristic spatial directions.

 \section {Clebsch Potentials}\label{SecClebsch}
 
 Some readers may be more familiar with the variational principle for perfect fluids in terms of Clebsch potentials. An irrotational velocity field can be described as the gradient of a scalar potential $T$. In discussing the vorticity of the fluid as in \eqref{fluidEq}, it is appropriate to consider $\mu u^\lambda$ rather than $u$ alone, and so we take $T$ to satisfy \[\mu u_\lambda \equiv \mu_\lambda= \partial_\lambda T.\]
 
 Then the fluid satisfies a variational principle with the Lagrangian equal to the pressure, which is taken to be a function of $\mu^2=g^{\kappa\lambda}\partial_\kappa T\partial_\lambda T$. Note that this is formally similar to the domain wall fluid discussed in the previous section. The only difference is that $\partial_\lambda T$ is here taken to be timelike rather than spacelike.
 
 If we wish to describe a fluid with nonvanishing vorticity we need to introduce additional scalar potentials. In a fluid with an entropy current in addition to a number density it is useful to consider four additional potentials as in a paper by Schutz \cite{Schutz:1970my}. We will delay the discussion of additional currents to the following section and consider a fluid with an equation of state depending on a single $n$. Then the vorticity is a simple bivector and we can describe the fluid with two additional scalar potentials $X$ and $Y$,
 \begin{align}
\mu = d T + X d Y. \label{clebsch}
 \end{align}
 As before, the pressure as a function of $\mu^2$ can be taken as the Lagrangian, and variations of $T,X$ and $Y$ lead to the correct fluid equations \cite{Schutz:1970my}.
 
 The vorticity takes the form of the flux tensor $\tF$,
 \begin{align}
 d\mu = dX\wedge dY \equiv \tF.
 \end{align}
Previously we were taking $\tF$ to describe the flux carried by strings in some underlying network, and the vorticity is indeed the flux carried by vortex lines in a superfluid. The superfluid may be described on a large scale such that the individual vortex lines are coarse-grained and the vorticity is a continuous tensor \cite{HV}\cite{BK}. One difference between this coarse-grained superfluid and an ordinary superfluid is that just as for a perfect string fluid, the thermodynamic quantities may depend on the magnitude of vorticity $\flux^2$ as well as the quantity $\mu^2$. Such a dependence appears already in the `vortex fibration model' of Carter and Langlois \cite{CarterLanglois}. In the remainder of this section we will show that their model of a superfluid at zero temperature follows from a simple modification of a perfect string fluid where the Lagrangian depends on $\mu$ and the scalar field $T$ (instead of $n$ and the earlier field $Z$).

First note that $T$ does not respect the same symmetries as $Z$. There is still symmetry under shifts of $T$ by a constant, which leads to a Noether current which will be identified as the ordinary fluid current $n$. But the quantity $\mu$ in \eqref{clebsch} is not preserved if we add an arbitrary function of $X,Y$ to $T$. Thus unlike the situation in \eqref{noether_perp}, $n$ is not in general orthogonal to $\tF$. In other words the vortex lines are not `frozen into' the fluid flow, in contrast to field lines in ideal magnetohydrodynamics.

However there is a symmetry under a simultaneous change in $X,Y$ and $T$. If $X,Y$ are changed by a symplectic transformation \eqref{symplectic}, \eqref{clebsch} will be preserved if $T$ is changed by
\begin{align}
\delta T = (H_X X - H)\delta t.\nonumber
\end{align}
And besides $\mu$ and $\tF$, the quantity
\begin{align}
\tilde{h}_{\lambda\mu\nu} \equiv (dX \wedge dY \wedge dT)_{\lambda\mu\nu} =  (\mu \wedge \tF)_{\lambda\mu\nu} = \mu\flux \tilde{w}_{\lambda\mu\nu}
\end{align}
also satisfies this symmetry. So in general we may take the Lagrangian  to also depend on
\begin{align}
h^2 \equiv \frac{1}{3!}\tilde{h}^{\lambda\mu\nu}\tilde{h}_{\lambda\mu\nu}.
\end{align}
in addition to $\flux^2$ and $\mu^2$. The duplicate notation $h$ is chosen in this context to agree with the notation for the helicity vector $h$ in the Carter-Langlois model \cite{CarterLanglois}. It is easy to see that a Lagrangian 
\begin{align}
\Lagr =  \Lagr\left (\frac{1}{2}\left (dX \wedge dY \right)^2, (dT + X dY)^2, -\frac{1}{3!}\left (dX \wedge dY \wedge dT \right )^2 \right).
\end{align}
leads to an energy-momentum tensor agreeing with that of Carter-Langlois. (Note that in Ref. \cite{CarterLanglois} the Lagrangian was denoted by $\Psi$).

The purpose of this section is rather to demonstrate that variation of $\Lagr$ by $X,Y$ and $T$ leads to an alternate variational principle to that of \cite{CarterLanglois}, which involves a Kalb-Ramond field rather than the scalar $T$ and requires an extra term in the Lagrangian to enforce a constraint. 

Defining the current $n$ and the antisymmetric tensor $\lambda$ through
\begin{align}
\delta \Lagr \equiv n^\rho \delta \mu_\rho + \frac{1}{2}\lambda^{\rho\sigma}\delta \tF_{\rho\sigma},
\end{align}
it is clear that the equation of motion resulting from a variation $\delta \phi$ is just the conservation
\begin{align}
\nabla_\rho n^\rho = 0.\nonumber
\end{align}

Variations by $\delta X$ and $\delta Y$ respectively lead to
\begin{align}
 n^\rho Y_{,\rho} - \nabla_{\rho}(\lambda^{\rho\sigma}Y_{,\sigma}) &= 0, \non
-\nabla_\rho(n^\rho X) - \nabla_{\sigma}(\lambda^{\rho\sigma}X_{,\rho}) &= 0. \nonumber
\end{align}
All these equations of motion for fields $X$, $Y$ and $Z$ can be combined to obtain the equation of motion of the Carter-Langlois model
\begin{align}
(n^\rho - \nabla_\sigma \lambda^{\sigma\rho})\tF_{\rho\tau} =0,
\end{align}
which reduces the ordinary equation of motion for a perfect fluid \eqref{fluidEq}, when the Lagrangian does not depend on $\tF$ (and thus $\lambda=0$).

\section{Additional Currents}\label{SecCurrents}
In the standard treatment of perfect fluids the equation of state is often taken to depend on both the number density $n$ and also a conserved entropy density $ n_s $. The problem of how to extend the variational principle to fluids with multiple constituents needs to be addressed. The most obvious solution is to implement the current $n_s$ by introducing an additional scalar field $ Z_s $ in the Lagrangian in the combination $ \tN_s = dX\wedge dY\wedge dZ_s $. Then the two currents, $n$ and $n_s$, may flow with different velocities, although both velocities will be confined to the same submanifolds.

If instead we wish for a current like entropy to flow with the same velocity $u$ as $n$, there are two valid options. First consider a modification similar to that taken in the diffeomorphic approach to ordinary fluids \cite{FluidReview}. The entropy per particle $ S $ is constant along the particle worldines, so it is a function $ S(X,Y,Z) $. The entropy density current is then
\begin{align}
\tN_s = S(X,Y,Z)\tN,
\end{align}
which is conserved by construction and points in the direction $u$. A Lagrangian depending on $ n_s^2 $ then can be varied by $ X,Y,Z $ (but not $ S $ itself) as in \eqref{fluidEq_string0}. The extra dependence on $n_s$ ultimately leads to an extra term in the equation of motion \eqref{fluidEq_string} of the form
\[ 2 n_s^\lambda \nabla_{[\kappa}\left (  \frac{\partial \Lagr}{\partial n_s} u_{\lambda]} \right ). \]
And this is just what is needed for conservation of the energy-momentum tensor \eqref{T} to hold if the equation of state also depends on $n_s$. 

In this approach even for an arbitrary number of additional currents, we do not introduce any extra degrees of freedom in the theory in the sense of extra fields having conjugate momenta. However the function $ S $, which physically depends on initial conditions, appears directly in the Lagrangian. This explicitly breaks the relabeling symmetries of $ X,Y,Z $.

Previously the Noether current associated to shifts in $Z$ was the dual current \eqref{dual_current}. It is indeed true that in the presence of additional currents (indexed by $a$) the dual current is not generalized to any gauge invariant Noether current. But a useful expression may still be derived from the $\delta Z$ equation of motion,
\begin{align}
n_a \nabla_\mu (\flux \mu^a w^\mu)=0.\label{dual_current_gen1}
\end{align} 
This can also be derived without making use of the variational principle by using energy-momentum conservation $w_\mu \nabla_\nu T^{\mu\nu}=0$, and the identity \eqref{carter_correspondence}.

An alternate approach to implementing additional currents flowing with the velocity has been previously suggested \cite{DHNS2012}. We may introduce an extra scalar field $\theta$, and allow the Lagrangian to also depend on $\theta$ through the combination
\begin{align}
y \equiv \frac{1}{n}\epsilon^{\kappa\lambda\mu\nu}X_{,\kappa}Y_{,\lambda}Z_{,\mu}\theta_{,\nu}.\nonumber
\end{align}

The equation of motion associated to $\delta \theta$ is
\begin{align}
\nabla_\mu(\Lagr_{,y} u^\mu)=0,
\end{align}
and thus $\Lagr_{,y}$ is interpreted as $n_s$. The quantity $y$ itself is equal to the chemical potential associated to $n_s$, and the Lagrangian in this case is the Legendre transform of $-\rho$ in the $n_s$ variable. In the specific case where $n_s$ is interpreted as entropy, $y$ is equal to the temperature $T$ and $\Lagr$ is the negative of the Helmholtz free energy. The field $\theta$ itself has appeared in the literature before as the quantity ``thermasy'' \cite{Schutz:1970my}.

This approach introduces the additional degree of freedom $\theta$, but maintains the relabeling symmetry in the Lagrangian. The Noether current associated to shifts in $Z$ now contains a gauge dependent term involving $\theta$
\begin{align}
\Pi_Z^\mu = \flux (\mu + ST)w^\mu - S F^{\mu\nu}\theta_{,\nu}.\label{dual_current_gen2}
\end{align}
However this dependence is eliminated upon taking the divergence
\begin{align}
\nabla_\mu \left(S F^{\mu\nu}\theta_{,\nu}\right) &= F^{\mu\nu}\theta_{,\nu}\nabla_\mu S\non&= \flux  Tw^\mu \nabla_\mu S,\nonumber
\end{align}
where the second line makes use of the vanishing of the derivative of $S$ in the $u$ direction. So the conservation of the current $\Pi_Z$ leads to the appropriate generalization  \eqref{dual_current_gen1} of the dual current conservation
\begin{align}
\nabla_\mu \Pi_Z^\mu = \nabla_\mu(\flux \mu w^\mu)+S\nabla_\mu(\flux T w^\mu)=0.
\end{align}

Finally note that both of these approaches to introducing extra currents can be easily generalized to introducing extra fluxes in the equation of state. For instance if we introduce dependence on the two fields $\theta^1,\theta^2$ in the combination
\[\upsilon \equiv \frac{1}{\flux}\epsilon^{\kappa\lambda\mu\nu}X_{,\kappa}Y_{,\lambda}\theta^1_{,\mu}\theta^2_{,\nu},\]
the two Noether currents associated with the new fields are equivalent to the conservation of a single antisymmetric tensor $ F_s = \Lagr_{,\upsilon} \Sigma$.  The conservation of $F_s$ and $F$ implies that $\Lagr_{,\upsilon}/\flux$ is constant on the worldsheets, and so we can represent it by a function $ S(X,Y) $. Alternately we could introduce $S(X,Y)$ directly in the Lagrangian through a dependence on $\flux_s^2 = S^2\flux^2$. And again the  Lagrangians for the two distinct approaches to introducing fluxes are simply related through Legendre transforms.

\section{Conclusion}\label{SecConclusion}

In this paper we developed a field theory description of various non-dissipative generalized fluids including pressureless dusts of topological defects, magnetohydrodynamics, and the Carter-Langlois model of a relativistic superfluid. \cite{CarterLanglois} The constructions are based on the variational principle which is discussed in detail in  Sec. \ref{SecVariational}. The effective degrees of freedom are the scalar fields $X, Y, Z$ with space-like gradients $dX, dY, dZ$ and a scalar field $T$ with a time-like gradient $dT$. In what follows we summarize different fluids discussed in the paper together with corresponding field theory Lagrangians $\Lagr$.

A perfect particle fluid may be described by an arbitrary function $\rho$ of the magnitude of $dX \wedge dY \wedge dZ$,
\begin{align}
\Lagr &= -\rho (n^2)\non
n^2 &\equiv-\frac{1}{3!} \left(dX \wedge dY \wedge dZ \right)^2,\nonumber
\end{align}
where $\rho$ is equal to the energy density, and $n$ the number density of the fluid.

The Stachel-Letelier string fluid describing a dust of Nambu-Goto strings may be described by a particular function of the magnitude of $dX \wedge dY$,
\begin{align}
\Lagr = - \mu_0 \sqrt{\frac{1}{2}\left (dX \wedge dY \right)^2  }\nonumber
\end{align}
where $\mu_0$ is string tension. 

This may be generalized to a {\it perfect string fluid} described by an arbitrary function $\rho$ of  both $dX \wedge dY$ and $dX \wedge dY \wedge dZ$, 
\begin{align}
\Lagr &=-\rho (n^2, \flux^2)\non
\flux^2 &\equiv\frac{1}{2} \left(dX \wedge dY  \right)^2,\nonumber
\end{align}
where $\flux$ has an interpretation as a flux carried by the strings.

This may be further generalized to describe fluids of higher dimensional branes as discussed in Sec.\eqref{SecDomainWalls}. In particular the Lagrangian
\begin{align}
\Lagr &=-\rho (n^2, \flux^2,\psi^2)\non
\psi^2&\equiv-\frac{1}{2}dX^2\nonumber
\end{align}
completely breaks the symmetry between $X,Y,Z$ and can be interpreted as a string fluid confined to a fluid of domain walls, where $\psi$ describes the local density of domain walls.

As described in Sec.\ref{SecCurrents}, dependence on the differentials of additional scalar fields wedged with the original fields $X,Y,Z$ generically leads to the presence of additional currents and fluxes. For instance the Lagrangian\[
\Lagr\left (\left (dX \wedge dY \right)^2, \left (dX \wedge dY \wedge dZ \right )^2 , \left ( dX \wedge dY \wedge Q \right)^2 \right)\]
describes a string fluid with the strings carrying an additional confined current or flux depending on whether $Q$ is a one-form or a two-form.

This variational principle may be applied to some familiar physical systems. In particular, ideal magnetohydrodynamics is equivalent to a string fluid with a particular form of the Lagrangian,
\begin{align}
\Lagr =-\rho_0(n^2) - \frac{1}{4}(dX \wedge dY)^2,
\end{align}
where $\rho_0(n^2)$ is energy density of the perfect fluid component of the plasma as a function of the number density, which again is expressed in terms of $dX \wedge dY \wedge dZ$. 

The Carter-Langlois model for a superfluid may be described in terms of the flux $\flux$ as before but the variational principle involves the Clebsch potential scalar field $T$ rather than $Z$, as described in Sec.\ref{SecClebsch}. The Lagrangian takes the general form 
\begin{align}
\Lagr\left (\left (dX \wedge dY \right)^2, (dT + X dY)^2,\left (dX \wedge dY \wedge dT \right )^2  \right).
\end{align}
Here $dT + X dY$ points in the direction of the velocity of the fluid $u$ and has magnitude equal to the chemical potential $\mu$. The extra dependence $dX \wedge dY \wedge dT$ may be interpreted as the dual of the helicity vector. 

The fluids discussed in this paper are isentropic, and another less obvious extension may be to introduce dissipation effects into string fluids using the Lagrangian framework developed here. A similar problem was studied in the context of black holes \cite{Goldberger} and more recently in context of particle fluids \cite{Nicolis}. Alternatively, an approach similar to that of Israel-Stewart \cite{israelStewart} has been used to introduce dissipative effects in string fluids and magnetohydrodynamics using the requirement of consistency with the second law of thermodynamics. \cite{Dissipative}


\begin{thebibliography}{10}

\bibitem{Vanchurin:2005yb} 
  V.~Vanchurin, K.~Olum and A.~Vilenkin,
``Cosmic string scaling in flat space,''
  Phys.\ Rev.\ D {\bf 72}, 063514 (2005)
  [gr-qc/0501040].

\bibitem{Schubring:2013qpa} 
  D.~Schubring and V.~Vanchurin,
  ``Fluid Mechanics of Strings,''
  Phys.\ Rev.\ D {\bf 88}, 083531 (2013)
  [arXiv:1305.6961 [hep-th]].
  
\bibitem{Vanchurin:2013tk} 
  V.~Vanchurin,
  ``Kinetic Theory and Hydrodynamics of Cosmic Strings,''
  Phys.\ Rev.\ D {\bf 87}, no. 6, 063508 (2013)
  [arXiv:1301.1973 [hep-th]].
  
\bibitem{StringFluid}
D.~Schubring and V.~Vanchurin,
``String fluid in local equilibrium,''
Phys.\ Rev.\ D {\bf 90}, 083516 (2014)
\bibitem{Dissipative}
D.~Schubring,
``Dissipative string fluids,''
Phys.\ Rev.\ D {\bf 91}, 043518 (2015)

\bibitem{Schutz:1970my} 
  B.~F.~Schutz,
  ``Perfect Fluids in General Relativity: Velocity Potentials and a Variational Principle,''
  Phys.\ Rev.\ D {\bf 2}, 2762 (1970).
  

\bibitem{DHNS2012}
  S.~Dubovsky, L.~Hui, A.~Nicolis and D.~T.~Son,
  ``Effective field theory for hydrodynamics: thermodynamics, and the derivative expansion,''
  Phys.\ Rev.\ D {\bf 85}, 085029 (2012)
  [arXiv:1107.0731 [hep-th]].


\bibitem{FluidReview}
N. Andersson, G.L. Comer,
``Relativistic Fluid Dynamics: Physics for Many Different Scales,''
Living Rev. Relativity {\bf 10}, (2007)


\bibitem{Kopczynski}
W. Kopczy\'{n}ski. ``A fluid of multidimensional objects,'' Phys.\ Rev.\ D {\bf 36}, 3582 (1987)

\bibitem{Stachel}
J.Stachel, ``Thickening the string. I. The string perfect dust,'' Phys.\ Rev.\ D {\bf 21}, 2171 (1980)

\bibitem{Letelier1979}
P. Letelier, ``Clouds of strings in general relativity,'' Phys.\ Rev.\ D {\bf 20}, 1294 (1979)

\bibitem{Letelier1983}
P. Letelier, ``String cosmologies,'' Phys.\ Rev.\ D {\bf 28}, 2414 (1983)

\bibitem{StringCosmology1990}
K. D. Krori, et al. ``Some exact solutions in string cosmology,'' General Relativity and Gravitation {\bf 22}, 123 (1990)

\bibitem{Wang2002}
X. Wang ``Exact solutions for string cosmology,'' Chinese Phys. Lett. {\bf 20} 615 (2003)

\bibitem{CarterBrane}
B. Carter. ``Dynamics of cosmic strings and other brane models,'' in Formation and Interactions of Topological Defects (NATO ASI B349), ed. R. Brandenberger, A.-C.
Davis, (Plenum, New York, 1995) 303-348. [arXiv : [hep-th/9611054]]

\bibitem{CarterDuality}
B.Carter. ``Duality relation between charged elastic stings and superconducting cosmic strings,'' Phys.\ Lett.\ {\bf B224}, 61 (1989)

\bibitem{CarterLanglois}
B.Carter and D.Langlois. ``Kalb-Ramond coupled vortex fibration model for relativistic superfluid dynamics,''  	Nucl.Phys. B{\bf 454} 402(1995) 

\bibitem{StringReview}
A. Vilenkin and E. P. S. Shellard,
Cosmic Strings and Other Topological Defects (Cambridge University Press, 2000).

\bibitem{Eckart1940}
C. Eckart. ``Relativistic Theory of the Simple Fluid,'' Phys. Rev. {\bf 58}, 919 (1940)

\bibitem{Harris1957}
E. G. Harris. ``Relativistic Magnetohydrodynamics,'' Phys. Rev. {\bf 108}, 1357 (1957)

\bibitem{Blackfold}
M. M. Caldarelli, R. Emparan, B. Van Pol. ``Higher-dimensional Rotating Charged Black Holes,'' JHEP {\bf 04} (2011) 013

\bibitem{Ciubotariu1973}
C. Ciubotariu and J. Gottlieb. ``The relativistic magnetohydrodynamics of a rotating plasma,'' Physica {\bf 63}, 393 (1973)

\bibitem{HV}
H. E. Hall and W. F. Vinen. ``The Theory of Mutual Friction in Uniformly Rotating
Helium II,'' Proc. Roy. Soc. A {\bf 238}, 215 (1956)
\bibitem{BK}
I. L. Bekharevich and I. M. Khalatnikov. ``Phenomenological derivation of the equations of vortex motion in He II,'' JETP {\bf 13}, 643 (1961)
 
  
 \bibitem{Goldberger} 
  W.~D.~Goldberger and I.~Z.~Rothstein,
  ``Dissipative effects in the worldline approach to black hole dynamics,''
  Phys.\ Rev.\ D {\bf 73}, 104030 (2006)
  [hep-th/0511133].
  
\bibitem{Nicolis} 
  S.~Endlich, A.~Nicolis, R.~A.~Porto and J.~Wang,
  ``Dissipation in the effective field theory for hydrodynamics: First order effects,''
  Phys.\ Rev.\ D {\bf 88}, 105001 (2013)
  [arXiv:1211.6461 [hep-th]].

 \bibitem{israelStewart}
 W. Israel and J. M. Stewart, Ann. Phys. {\bf 118}, 341 (1979)
 
  \end{thebibliography}
\end{document}